\documentclass{article}
\begin{document}

\title{Origins of Modern Data Analysis Linked to the
Beginnings and Early Development of Computer Science
and Information Engineering}

\author{Fionn Murtagh  \\
Science Foundation Ireland, Wilton Place, Dublin 2, Ireland, and \\
Department of Computer Science, Royal Holloway University of London, \\
Egham TW20 0EX, England \\
Email fmurtagh@acm.org}
\maketitle

\begin{abstract}
The history of data analysis that is addressed here is underpinned
by two themes, -- those of tabular data analysis, and the analysis 
of collected heterogeneous data.  ``Exploratory data analysis'' 
is taken as the heuristic approach that begins with data and 
information and seeks underlying explanation for what is 
observed or measured.  I also cover some of the evolving 
context of research and applications, including scholarly 
publishing, technology 
transfer and the economic relationship of the university to 
society.
\end{abstract}

\section{Data Analysis as the 
Understanding of Information and Not Just Predicting Outcomes}

\subsection{Mathematical Analysis of Data}

The mathematical treatment of data and information has been recognized
since time immemorial.  Galileo \cite{galileo} expressed it in this way:
``Philosophy is written in this immense book that stands ever open before
our eyes (I speak of the Universe), but it cannot be read if one does
not first learn the language and recognize the characters in which it
is written.  It is written in mathematical language, and the characters
are triangles, circles, and other geometrical figures, without the
means of which it is humanly impossible to understand a word; without
these philosophy is confused, wandering in a dark labyrinth.''
Plato is reputed to have had the phrase ``Let no-one ignorant of 
geometry enter'' at the entrance to his Academy, the school he founded
in Athens \cite{suzanne}.  

\subsection{Collecting Data}

Large scale collection of data supports the analysis of data. 
Such collection is facilitated greatly by modern computer science
and information engineering hardware, middleware and software.  
Large scale data collection in its own right does not necessaraily 
lead to successful exploitation, as I will show with two examples, 
the VDI technical lexicon and the Carte du Ciel object inventory.  

Walter Banjamin, 1882--1940, social and media (including photography and film)
technology critic, noted the following, relating to 
engineering  \cite{benjamin}.  
``Around 1900, the Verband Deutscher Ingenieure [German Engineers' 
Association] set to work on a comprehensive technical lexicon.
Within three years, index cards for more than three-and-a-half 
million words had been collected.  But `in 1907 the association's
managing committee calculated that, with the present number of 
personnel, it would take forty years to get the manuscript of 
the technical lexicon ready for printing.  The work was abandoned
after it had swallowed up half a million marks' \cite{wuester}. It
had become apparent that a technical dictionary should be 
structured in terms of its subject matter, arranged systematically.
An alphabetical sequence was obsolete.'' 
The ``Carte du Ciel'' (sky map) project was, in a way, similar.
It was started in 1887 by Paris Observatory and the aim was to map 
the entire sky down to the 11th or 12th magnitude.  It was 
planned as a collective and laborious study of photographic plates that
would take up to 15 years.  The work was not completed.  There was a 
widespread view \cite{jones}
that such manual work led to French, and European, astronomy falling 
behind other work elsewhere that was driven by instrumentation and 
new observing methods.  

Over the past centuries, 
the mathematical treatment of data and information became nowhere more 
core than in physics.  As Stephenson notes in his novel
(\cite{stephenson}, p. 689) -- a point discussed by other authors 
elsewhere also -- Isaac Newton's
``Principia Mathematica might never have come about had Nature not 
sent a spate of comets our way in the 1680s, and so arranged their
trajectories that we could make telling observations.''
Indeed the difficulties of experimentally verifying the theory of
superstrings has led to considerable recent debate (e.g., recent 
books by P. Woit, 
{\em Not Even Wrong}; and L. Smolin, {\em The Trouble with Physics}). 

\subsection{Data Analysis and Understanding}

I will move now to some observations 
related to the epistemology facets of data analysis 
and data mining.  I use the cases of G\"odel and of Benz\'ecri 
to show how earlier thinkers and scientists were 
aware of the thorny issues in moving from observed or measured data to
the explanatory factors that underlie the phenomena associated with 
the data.   

Logician Kurt G\"odel, 1906--1978,  had this critique to make of 
physics: ``physics ... combines concepts without
analyzing them'' (p. 170, \cite{yourgrau}). 
In a 1961 perspective he was able to claim: 
``... in physics ... the possibility of knowledge of objectivizable
states of affairs is denied, and it is asserted that we must be content
to predict the results of observations.  This is really the end of
all theoretical science in the usual sense.'' (p. 140, \cite{yourgrau}.)  



This was a view that was in many ways shared by the data analysis
perspective espoused by data analyst and theorist, Jean-Paul 
Benz\'ecri \cite{jpb0}: 
``... high energy theoretical
physics progresses, mainly, by constituting
corpora of rare phenomena among immense sets of ordinary cases.  The simple
observation of one of these ordinary cases requires detection apparatus
based on millions of small elementary detectors. ...

Practitioners walk straight into the analyses, and transformations ... 
without knowing what they are looking for.  

What is needed, and what I am proud about having somewhat succeeded in
doing, is to see what is relevant in the objects studied, ...''

Philosopher Alain 
Badiou \cite{badioux} (p. 79) echoes some of these perspectives on 
blindly walking into the analysis task: 
``empirical prodigality becomes something like an arbitrary and
sterile burden.  The problem ends up being replaced by {\em verification}
pure and simple.''

The debate on the role of data analysis is not abating.  In 
data analysis in neuroscience (see \cite{miller} or, as \cite{hauser}
states in regard to the neurogenetic basis of language, ``studies 
using brain imaging must acknowledge that localization of function does 
not provide explanatory power for the linguist attempting to uncover
principles underlying the speaker's knowledge of language''), 
the issue of what 
explains the data analyzed is very often unresolved.  This is notwithstanding 
solid machine or computer learning 
progress in being able to map characteristics in the data 
onto outcomes.   

In this short discussion of data analysis, I am seeking solely to 
demarcate how I understand data analysis in this article. 
In general terms this is also what goes under the terms of:
{\em analyse des donn\'ees} in the French tradition; data mining; 
and unsupervised classification.  The latter is the term used
in the pattern recognition literature, and it can be counterposed
to supervised classification, or machine learning, or discriminant
analysis. 

\section{Beyond Data Tables: Origins of Data Analysis}
\label{sect2}

\subsection{Benz\'ecri's Data Analysis Project}

For a few reasons I will begin with a focus on correspondence analysis.
Following the best part of two years that I spent using multidimensional
scaling and other methods in educational research, I started on 
a doctoral program in Benz\'ecri's lab in 1978.  I was very  
impressed by the cohesiveness of theoretical underpinning and 
breadth of applications there. 
Rather than an ad hoc application of an analysis method to a
given problem,
there was instead a focused and integrated view of theory and
practice.  What I found was far from being a bag of analytical 
tricks or curiosities.  Instead the physics or psyche or social 
process lying behind the data was given its due. 
Benz\'ecri's early data analysis motivation sprung 
from text and document analysis.  I will return to these areas
in section \ref{sect43} below.  

The correspondence analysis and ``aids to                      
interpretation'' research programs as well as software programs 
were developed and deployed on a broad scale 
in Benz\'ecri's laboratory at the Universit\'e Pierre et Marie 
Curie, Paris 6, through the 1970s,
1980s and 1990s.
The hierarchical clustering programs distill the best of the
reciprocal nearest neighbors algorithm that was published
in the early 1980s in 
Benz\'ecri's journal, {\em Les Cahiers
de l'Analyse des Donn\'ees} (``Journal of Data Analysis'')
and have not been bettered since then.  (See also section 
\ref{sectpub} below.)
Much of the development work in this
framework, ranging from Einstein tensor notation through to
the myriad application studies published in the journal
{\em Les Cahiers de l'Analyse des Donn\'ees}, were in an 
advanced state of development 
by the time of my arrival in Paris in late 1978 to start on a
doctoral program.

A little book published in 1982, {\em Histoire et Pr\'ehistoire de             
l'Analyse des Donn\'ees} \cite{benzecri82}, offers
insights into multivariate data analysis or multidimensional
statistics.  It was written in the spring
of 1975, circularized internally, published chapter-wise
in {\em Les Cahiers de l'Analyse des Donn\'ees}, 
before taking book form.
It begins with a theme that echoes widely in Benz\'ecri's
writings: namely that the advent of computers overturned statistics
as understood up until then, and that the suppositions and premises
of statistics had to be rethought in the light of computers.  From
probability theory, data analysis inherits inspiration but not
methods: statistics is not, and cannot be, probability alone.
Probability is concerned with infinite sets, while data analysis
only touches on infinite sets in the far more finite world expressed by
such a typical problem as discovering the system of relationships
between rows and columns of a rectangular data table.

\subsection{Tabular Data}

With a computational infrastructure, the analysis of data tables 
has come into its own.  Antecedents clearly go back much 
further.  Therefore let  
me place the origins of data analysis in an unorthodox setting.
Clark \cite{clark} cites Foucault 
\cite{foucault} approvingly: ``The constitution of 
tables was one of the great problems of scientific, political and economic
technology in the eighteenth century ... The table of the eighteenth 
century was at once a technique of power and a procedure of knowledge''.  

If tabular data led to data analysis, then it can also be pointed out that
tabular data -- in another line of evolution -- led to the computer.
Charles Babbage, 1791--1871, is generally avowed to be a father of the 
computer \cite{swade}.  
Babbage's early (mechanical) versions of computers, his Difference Engine
and Analytical Engine, were designed with tabular data processing 
in view, for example generating tables ranging from logarithms to longitude.  
``The need for tables and the reliance placed on them became especially 
acute during the first half of the nineteenth century, which witnessed 
a ferment of scientific invention and unprecedented engineering ambition --
bridges, railways, shipbuilding, construction and architecture.  ...
There was one need for tables that was paramount -- navigation.  ...
The problem was that tables were riddled with errors.'' \cite{swade}.
The computer was called for to avoid these errors.  

\subsection{Algorithmic and Computational Data Analysis}

In discussing R.A.\ Fisher (English statistician, 1890--1962), 
Benz\'ecri \cite{benzecri82} acknowledges that 
Fisher in fact developed the basic
equations of correspondence analysis but without of course a desire to
do other than address the discrimination problem. 
Discriminant analysis, or supervised classification, took off in a major
way with the availability of computing infrastructure.  The
availability of such methods in turn motivated a great deal of work
in pattern recognition and machine learning.
It is to be noted that computer-based
analysis leads to a change of perspective
with options now available that were not
heretofore.  Fisher's brilliant approaches implicitly assume that variables
are well known, and that relations between variables are strong.  In
many other fields, a more exploratory and less precise observational
reality awaits the analyst.

With computers came pattern recognition and, at the start, neural networks.
A conference held in Honolulu in 1964 on {\em Methodologies of Pattern         
Recognition}, that was attended by Benz\'ecri,
 cited Rosenblatt's perceptron work many times (albeit his work
was cited
but not the perceptron as such).  Frank Rosenblatt (1928--1971) was 
a pioneer of neural networks, including the perceptron and 
neuromimetic computing which he developed in the 1950s.  
Early neural network research was simply
what became known later as discriminant analysis.  The problem of
discriminant analysis, however, is insoluble if the characterization of
observations and their measurements are not appropriate.  This leads
ineluctably to the importance of the data coding issue for any type of
data analysis.

Psychometrics made multidimensional or multivariate data analysis what it
has now become, namely, ``search by induction                                  
of the hidden dimensions that are defined by combinations of primary           
measures''. 
Psychometrics is a response to the problem of exploring areas where
immediate physical measurement is not possible, e.g.\ intelligence,
memory, imagination, patience.  Hence a statistical construction is used
in such cases (``even if numbers can never quantify the soul!''
\cite{benzecri82}).

While it is now part of the history of data analysis and statistics that
around the start of the 20th century interest came about in human
intelligence, and an underlying measure of intelligence, the intelligence
quotient (IQ), there is a further link drawn by Benz\'ecri \cite{benzecri82}
in 
tracing an astronomical origin to psychometrics.  Psychophysics, as also 
many other analysis frameworks such as the method of
least squares, was developed in no small way by astronomers: the desire to
penetrate the skies led too to study of the scope and limits of human
perception, and hence psychometrics. 

Around the mid-1960s Benz\'ecri began a correspondence with Roger 
N.\ Shepard
which resulted in a visit to Bell Labs.  Shepard (``a statistician only        
in order to serve psychology, and a psychologist out of love for               
philosophy'') and J.\ Douglas 
Carroll (who ``joyfully used all his ingenuity --         
which was large indeed -- to move data around in the computer like one         
would move perls in a kaleidoscope'') had developed proximity analysis,
serving as a lynchpin of multidimensional scaling.

\subsection{A Data Analysis Platform}

The term ``correspondence analysis'' was first proposed in the
fall of 1962.  
The first presentation under this title was made by J.-P. Benz\'ecri at the
Coll\`ege de France in a course in the winter of 1963.

By the late 1970s
what correspondence analysis had become was not limited to the extraction
of factors from any table of positive values.  It also catered for data
preparation; rules such as coding using complete disjunctive form; tools
for critiquing the validity of results principally through calculations
of contribution; provision of effective procedures for discrimination and
regression; and harmonious linkage with cluster analysis.  Thus a
unified approach was developed, for which the formalism remained quite
simple, but for which deep integration of ideas was achieved with diverse
problems.  Many of the latter originally appeared from different sources,
and some went back in time by many decades.

Two explanations are
proposed in \cite{benzecri82} 
for the success of correspondence analysis.  Firstly, the
principle of distributional equivalence allows a table of positive values
to be given a mathematical structure that compensates, as far as possible,
for arbitrariness in the choice of weighting and subdivision of categories.
Secondly, a great number of data analysts, working in very different
application fields, found available a unified processing framework, and a
single software package.
Correspondence analysis was considered as a standard, unifying and
integrated analysis framework -- a platform.

\subsection{Origins in Linguistic Data Analysis}

Correspondence analysis was initially proposed as an inductive method
for analyzing linguistic data.  From a philosophy standpoint,
correspondence analysis simultaneously processes large sets of
facts, and contrasts them  
in order to discover
global order; and therefore it has more to do with synthesis
(etymologically, to synthesize means to put together) and induction.
On the other hand, analysis and deduction  (viz., to distinguish the
elements of a whole; and to consider the properties of the possible
combinations of these elements) have become the watchwords of data
interpretation.  It has become traditional
now to speak of data analysis and correspondence analysis, and not
``data synthesis'' or ``correspondence synthesis''.      

The structural linguist Noam Chomsky, in the little 
volume, {\em Syntactic Structures} \cite{chomsky}, held that
there could not be a systematic procedure for determining the grammar
of a language, or more generally linguistic structures, based on a set of
data such as that of a text repository or corpus.  Thus, for
Chomsky, linguistics cannot be inductive (i.e., linguistics cannot
 construct itself using
a method, explicitly formulated, from the facts to the laws that
govern these facts); instead linguistics has to be deductive (in the
sense of starting from axioms, and then deriving models of real
languages).

Benz\'ecri did not like this approach.  He found it
 idealist, in that
it tends to separate the actions of the mind from the facts that are
the inspiration for the mind and the object of the mind.  At that time
there was not available an effective algorithm to take ten
thousand pages of text
from a language to a syntax, with the additional purpose of
yielding semantics.  
But now, with the advances in our computing infrastructure,
statistics offers the linguist an effective
inductive method for usefully processing data tables that one can
immediately collect, with -- on the horizon -- the ambitious layering
of successive research that will not leave anything in the shade --
from form, meaning or style.

This then is how data analysis is
feasible and practical in a world fueled by computing capability:
``We call the distribution of a word the set                                   
of its possible environments.''  
In the background there is  a consideration that Laplace noted:
 a well-constructed language automatically
leads to the truth, since faults in reasoning are shown up as faults
in syntax.  Dijkstra, Wirth, Hoare and the other pioneering 
computer scientists who developed the bases of 
programming languages that we use today, could not have expressed this 
better.  Indeed, Dijkstra's view was that 
``the programmer should let correctness proof and program grow hand in hand''
\cite{dijkstra}.  

\subsection{Information Fusion}

From 1950 onwards, statistical tests became very popular,
to verify or to protect the acceptability of a hypothesis (or of a
model) proposed a priori.  On the other hand correspondence analysis refers
from the outset to the hypothesis of independence of observations
(usually rows) $I$ and attributes (usually columns) $J$ but
aims only at exploring the extent to which this is not verified:
hence the spatial representation of uneven affinities between
the two sets.
Correspondence analysis looks for typical models that are achieved
a posteriori and not a priori.  This is following the application of
mutual processing of all data tables, without restrictive hypotheses.
Thus the aim is the inductive conjugating of models.  

\subsection{Benz\'ecri's and Hayashi's Shared Vision of Science}

If Benz\'ecri was enormously influential in France in drawing out
the lessons of data analysis being brought into a computer-supported
 age, in Japan Chikio Hayashi played no less a role.  Hayashi (1918--2002)
led areas that included public opinion research and statistical
mathematics, and was first president of the Behaviormetric Society of Japan.

In \cite{hayashi} Hayashi's data analysis approach is set out very clearly.

Firstly, what Hayashi referred to as ``quantification'' was the
scene setting or data encoding and representation 
forming the basis of subsequent
decision making.  He introduced therefore \cite{hayashi}
``methods of quantification of qualitative data in multidimensional analysis 
and especially how to quantify qualitative patterns to secure the 
maximum success rate of prediction of phenomena from the statistical 
point of view''.  So, firstly data, secondly method, and thirdly
decision making are inextricably linked.

Next comes the role of data selection, weighting, decorrelation,
low dimensionality selection and related aspects of the analysis,
and classification.  ``The important problem in multidimensional 
analysis is to devise the methods of the quantification of complex 
phenomena (intercorrelated behaviour patterns of units in dynamic 
environments) and then the methods of classification. Quantification 
means that the patterns are categorized and given numerical values 
in order that the patterns may be able to be treated as several 
indices, and classification means prediction of phenomena.''
In fact the very aim of factor analysis type analyses, including
correspondence analysis, is to
prepare the way for classification: ``The aim 
of multidimensional quantification is to make numerical representation 
of intercorrelated patterns synthetically to maximize the 
efficiency of classification, i.e. the success rate of prediction.''
Factorial methods are insufficient in their
own right, maybe leading just to display of data:
``Quantification does not mean finding numerical values but giving
them patterns on the operational point of view in a proper sense.
In this sense, quantification has not absolute meaning but relative
meaning to our purpose.''

This became very much the approach of Benz\'ecri too.  Note that
Hayashi's perspectives as described above date from 1954.  In
\cite{jpbavenir}, a contribution to the journal Behaviormetrika
that was invited by Hayashi, Benz\'ecri draws the following
conclusions on data analysis: ``In data analysis numerous disciplines
have to collaborate.  The role of mathematics, although essential,
remains modest in the sense that classical theorems are used almost
exclusively, or elementary demonstration techniques.  But it is
necessary that certain abstract conceptions penetrate the spirit of
the users, who are the specialists collecting the data and having to
orientate the analysis in accordance with the problems that are
fundamental to their particular science.''  This aspect of integral
linkage of disciplines is as aspect that I will return to in the
Conclusions.

Benz\'ecri \cite{jpbavenir} develops the implications of this.
The advance of compute capability (remember that this article
was published in 1983) ``requires that Data
Analysis [in upper case indicating the particular sense of data
analysis as -- in Hayashi's terms and equally the spirit of Benz\'ecri's
work -- quantification and classification] project ahead of the concrete
work, the indispensable source of inspiration, a vision of science.''
Benz\'ecri as well as Hayashi developed data analysis as 
projecting a vision of science.  

He continues: ``This vision is philosophical: it is not a matter of
translating directly in mathematical terms the system of concepts of a
particular discipline but of linking these concepts in the equations
of a model.  Nor is it a matter of accepting the data such as they are
revealed, but instead of elaborating them in a deep-going synthesis
which allows new entities to be discovered and simple relationships
between these new entities.''

Finally, the overall domain of application of data analysis is
characterized as follows: ``Through differential calculus, experimental
situations that are admirably dissected into simple components
were translated into so many fundamental laws.  We believe that it is
reserved for Data Analysis to express adequately the laws of that which,
complex by nature (living being, social body, ecosystem), cannot be
dissected without losing its very nature.''

While Hayashi and Benz\'ecri shared a vision of science, they also
shared greatly a view of methodology to be applied.  In a 1952
publication \cite{hayashi0} Hayashi referred to ``the problem of 
classification by quantification method'' which is not direct and
immediate clustering of data, but rather a careful combination of
numerical encoding and representation of data as a basis for the
clustering.  Hayashi's aim was to discuss:  ``(1) the methods of 
quantification of qualitative statistical data obtained by our measurements 
and observations ...; (2) ... the patterns of behaviour must be 
represented by some numerical values; (3) ...  effective grouping 
is required.''  Data analysis methods are not applied in isolation,
therefore.  In \cite{jpbavenir} Benz\'ecri referred to correspondence
analysis and hierarchical clustering, and indeed discriminant analysis
(``so as not to be illusory, a discriminant procedure has to be 
applied using a first set of cases -- the base set -- 
and then trialled on other cases -- the test set'').

In \cite{jpbavenir} Benz\'ecri refers, just a little, to the
breakthrough results achieved in hierarchical clustering algorithms
around this time, and described in the work of Juan \cite{juan1,juan2}.
These algorithmic results on hierarchical clustering were furthered
in the following year by the work of de Rham \cite{derham}.  As computational
results they have not been bettered since and still represent the
state of the art in algorithms for this family of classification
method.  In \cite{mur83a,mur84,murtagh85} I presented surveys of these
algorithms relative to other fast algorithms for particular
hierarchical clustering methods, and my software code was used in the CLUSTAN
and R packages.  More software code is available at \cite{mursw}.
In \cite{mur83b} I showed how, in practice, even
more efficient algorithms can be easily designed.

\section{The Changing University: Academe, Commercialization and Industry}

\subsection{The Growing Partnership of University and Industry}

It is generally considered that a major milestone -- perhaps the 
most important event -- in propelling the university into the modern
age was the Baye-Dole Act, brought into legislation in the United
States, in 1980.  It was a radical change in public polity perspective 
on the university and on research.  Prior to then, the research role 
of universities was a public role, and research results were to be 
passed on to industry.  Intellectual property was owned, prior to 
1980, by the public purse, as embodied in the US Government.  The Baye-Dole
Act allowed universities to own intellectual property, to license it,
and to otherwise exploit it as they saw fit.  The university 
became a close partner of industry.  It was a change in 
legislation and in perception that echoed around the planet.  

\subsection{1968 in France: Role in Bringing the University Closer to
Industry}
\label{jpb68}

The rise of the partnership between the university and industry, 
between academe and commercialization, that is now so integral,
everywhere, had other facets too.  Benz\'ecri's reflections 
\cite{benzecri07} in this regard are of interest.  In fact, these
reflections throw a somewhat different light on the 1968 period of 
significant student and general social unrest.  

Benz\'ecri \cite{benzecri07} paints the following picture. 

``Forty years ago a memorable academic year started, the ravages
of which are often deplored but also I must confess to having  
been the happy beneficiary.

Charged by Prof.\ Daniel Dugu\'e with the teaching of the 
Dipl\^ome d'Etudes Approfondies (DEA) de Statistique -- the Advanced 
Studies degree in Statistics, constituting the first part of a 
doctorate, I chose Data Analysis as the theme of the course.
Prof.\ Dugu\'e laughed and said this covered all of statistics!
Under this broad banner, I intended to carry out lots of 
correspondence analyses.  With such analyses, thanks to the 
patience of Brigitte Cordier, working on an IBM 1620 -- a 
pocket calculator today but one that the Dean Yves Martin 
provided for the price of a chateau! -- I was able, in Rennes,
to aim at conquering linguistics, economics, and other fields.

To analyze, data were necessary.  I was resolved to send the
students of the DEA  to collect sheafs of these precious flowers.

From my first class, I announced that the students should 
undertake internships.  But this call, repeated from week to week,
had no response.  The students thought that, if they survived a 
written exam and an oral then no-one could ask them for more.  
Other than by cramming, they would have nothing to gain from the 
novelty of an excursion into practical things.  Moreover, even if 
they accepted or even were enticed by my project then who would 
host their internship?  

The pretty month of May 1968 would change all that!

Living in Orl\'eans, I did not hear the shouting from Lutetia 
[i.e.\ from Paris and also from the university; Lutetia, or 
Lut\`ece, is the name of a town pre-existing Paris and the 
Ar\`enes de Lut\`ece
is a public park very close to Universit\'e Pierre et Marie Curie, 
Paris 6]  but 
only distant echoes.  Finally, in September, both boss and 
students had to resign themselves to each take back their role,
each cautiously but also brazenly!

Since the month of May, the university was stigmatized as being 
arthritic, not offering a preparation for life.  Not I but others
had extolled internships.  Machine-like of course the students came 
to tell me that they wanted to carry out internships.  I had triumphed!
Yes, yes, they stammered in confession, you had told us. 

It was just so too for those who in November 1967 had refused any 
intern now in September 1968 were keen to save their native land 
by pampering young people.  The way opened up for correspondence 
analysis, a methodology that in practice was very soon associated 
with hierarchical clustering.''  

\subsection{The Changed Nature of the PhD}

It is a source of some pride for me to be able to 
trace back through my doctoral lineage as follows \cite{mathgen}.  
My 
PhD advisor was Jean-Paul Benz\'ecri. He studied with Henri 
Cartan (of Bourbaki: see \cite{aczel} for a survey). 
Tracing back through advisors or mentors I have: \'Emile Borel and 
Henri Lebesgue; Simeon Poisson (advisor also to Gustav Dirichlet and 
Joseph Liouville); Joseph Lagrange (advisor also of Jean-Baptiste 
Fourier); Leonhard Euler; Johann Bernoulli; Jacob Bernoulli; and 
Gottfried Leibniz.

The PhD degree, including the title, the dissertation and the
evaluation framework
as a work of research (the ``rite of passage'')
came about in the German lands between the 1770s and the 1830s.  Clark
\cite{clark} finds it surprising that it survived the disrepute
associated with all
academic qualifications in the turmoil of the late 18th century.  In the
United States, the first PhD was awarded by Yale University in 1861.  In the
UK, the University of London introduced the degree between 1857 and 1860.
Cambridge University awarded the DPhil or PhD from 1882, and Oxford
University only from 1917.

A quite remarkable feature of the modern period 
is how spectacular the growth of PhD numbers has now become.
In \cite{murtagh08}, I discuss how in the US, to take one example, 
in Computer Science and Engineering, the number of PhDs awarded has
doubled in the three years to 2008.  
Internationally this evolution holds too.
For example, Ireland is pursuing a doubling of PhD output up
to 2013.

Concomitant with numbers of PhDs, the very
structure of the PhD is changing in many countries outside North
America.  There is a strong movement away from the traditional
German ``master/apprentice'' model, towards instead a ``professional''
qualification.  This move is seen often as towards the US model.
In Ireland there is a strong move to reform the PhD towards what is
termed a
``structured PhD''.  This involves a change from the apprenticeship model
consisting
of lone or small groups of students over three years in one university
department to a
new model incorporating elements of the apprenticeship model centered around
groups of students possibly in multiple universities where generic and
transferable
skills (including entrepreneurial) can be embedded in education and
training over four
years.
Unlike in most of Europe, Germany is retaining a traditional
``master/apprentice'' model.

Numbers of PhDs are dramatically up, and in many countries there is a
major restructuring underway of the PhD work content and even timeline.
In tandem with this, in North America
the majority of PhDs in such areas as computer science and computer 
engineering now move directly into industry when they graduate.  
This trend goes
hand in hand with the move from an apprenticeship for a career in
academe to, instead, a professional qualification for a career in
business or industry.

In quite a few respects Benz\'ecri's lab was akin to what is now widely
targeted in terms of courses and  large scale production of PhDs.  I 
recall at the end of the 1970s  how there were about 75 students 
{\em en th\`ese} -- working on their dissertations -- 
and 75 or so attending courses in
the first year of the doctoral program leading to the DEA, Dipl\^ome 
d'\'Etudes Approfondies, qualification.  Not atypically at this 
time in terms of
industrial outreach, I carried out 
a study for a company CIMSA, Compagnie d'Informatique Militaire, Spatiale
et A\'eronautique (a subsidiary of Thomson) on 
future computing needs; and my thesis was in conjunction with the
Bureau de Recherches G\'eologiques et Mini\`eres, Orl\'eans, 
the national geological research and development agency.  

\subsection{Changing Citation and Other Aspects of Scholarly 
Publication Practice}
\label{sectpub}

In my time in Benz\'ecri's lab, I developed a view of
citation practice, relevant for mathematically-based PhD research in
the French tradition, and this was as follows: a good introduction
in a PhD dissertation
in a mathematical domain would lay down a firm foundation in
terms of lemmas, theorems and corollaries with derivation of results.
However there was not a great deal of citing.  Showing that one
had assimilated very well the content was what counted, and
not reeling off who had done what and when.  On the other hand, it
seemed to me to be relatively clear around 1980 that in general a
PhD in the ``Anglo-Saxon countries'' would start with an
overview chapter containing plenty of citations to the relevant
literature (or literatures).  This different tradition aimed at
highlighting what one was contributing in the dissertation, after
first laying out the basis on which this contribution was built.

Perhaps the divide was indicative just of the strong mathematical
tradition in the French university system.
 However more broadly speaking citation practices have
changed enormously over the past few decades.  I will dwell a
little on these changes now.

In a recent (mid-2008) review of citation statistics, Adler 
et al.\ \cite{adler}
note that in mathematics and computer science, a published article is
cited on average less than once; in chemistry and physics, an article
is cited on average about three times; it is just a little
higher in clinical medicine; and in the life sciences a
published article is on average cited more than six times.
It is small wonder therefore that in recent times (2008) the NIH (National
Institutes of Health, in the US) has been a key front-runner in
pushing Open Access developments -- i.e.\ mandatory depositing of
article postprints upon publication or by an agreed date following
publication.  The NIH Open Access mandate was indeed legislatively enacted
\cite{nihoa}.

In his use of {\em Les Cahiers de l'Analyse des Donn\'ees}, a journal
published by Dunod and running with 4 issues each year over 21
years up to 1997, Benz\'ecri focused and indeed concentrated
the work of his lab.  Nowadays a lab- or even institute-based journal
appears unusual even if it certainly testifies to a wide range of
applications and activities.  It was not always so.  Consider e.g.\
the {\em Journal f\"ur die reine und angewandte Mathematik}, referred to
as {\em Crelle's Journal} in an earlier age of mathematics when August
Leopold Crelle had founded it and edited it up to his death in 1855.
Or consider, closer to
Benz\'ecri's lab, the {\em Annales de l'ISUP}, ISUP being the Institut
Statistique de l'Universit\'e de Paris.

It is of interest to dwell here on just what scientific, or scholarly,
publication is, given the possible insight from the past into current
debates on Open Access, and citation-based performance and resource-allocation
models in national research support systems.

As is well known what are
commonly regarded as the first scientific journals came about in 1665.
These were the Philosophical Transactions of the Royal Society of London
early in that year and the Journal des S\c{c}avants in Paris a little
later.  Gu\'edon \cite{guedon} contrasts them, finding that
``the Parisian publication followed novelty while the 
London journal was helping to validate originality''.  The Philosophical
Transactions was established and edited by Henry Oldenburg (c.\ 1619 to 1677),
Secretary of the Royal Society.  This journal
``aimed at creating a public record of original contributions  
to knowledge''.  Its primary function was not (as such) general
communication between peers, nor dissemination to non-scientists,
but instead ``a public registry of discoveries''.
This first scientific journal was a means for creating intellectual property.
Journals originating in Oldenburg's prototypical scientific --
indeed scholarly -- journal are to be seen ``as registers of 
intellectual property whose functions are close to that of a land 
register''.  

Noting parallels with the modern web age, Gu\'edon \cite{guedon} sees 
how in the 17th century, ``the roles of writers, printers, and bookstore 
owners, as well as their boundaries, were still contentious topics.''
The stationers sought to establish their claim, just like a claim to 
landed property.  By defining authorship in the writing activity, 
and simultaneously the 
intellectual property of that author, the way was open to the 
stationer as early publisher to have the right to use this 
property analogous to landed property.  
Johns \cite{johns} points to how
suspicion and mistrust accompanied early publishing so that Oldenburg
was also targeting an ``innovative use of print technology''.
Gu\'edon \cite{guedon} finds: ``The design of a scientific
periodical, far from primarily
aiming at disseminating knowledge, really seeks to reinforce property
rights over ideas; intellectual property and authors were not legal
concepts designed to protect writers -- they were invented for the
printers' or stationers' benefits.'' 
Let me temper this to note how important intellectual property
over ideas is additionally in terms of motivation of scholars in 
subsequent times.  

The context as much as the author-scientist
led to this particular form of intellectual property.  What I find
convincing enough in the role of the printer or stationer is that
the article, or collection of articles in a journal, or other forms of
printed product (pamphlet, treatise), became the most important paradigm.
Other possible forms did not.  Examples could include: the experiment;
or the table of experimental data; or catalogs or inventories. 
Note that the latter have become extremely important in, e.g.,
observational data based sciences such as astronomy, or the processed
or derived data based life sciences.  What is interesting is that the
publication remains the really dominant form of research output.  

Coming now to authorship, there has been an overall shift towards
teamwork in authorship, clearly enough led by the life sciences and 
by ``big science''.  In the highly cited journal, {\em Nature}, 
it has been noted \cite{whitfield} that ``almost all original research
papers have multiple authors''.  Furthermore (in 2008),
``So far this year ... {\em Nature} has published only six 
single-author papers, out of a total of some 700''.  
What is however very clear is that mathematics or statistics
or related methodology work rarely ever appears in {\em Nature}.
While social networks of scientists have become very important,
notes Whitfield, nonetheless there is room still for a 
counter-current in scholarly activity:
``... however finely honed scientists' team-building strategies 
become, there will always be room for the solo effort. In 1963, Derek 
de Solla Price, the father of authorship-network studies, noted that 
if the trends of that time persisted, single-author papers in chemistry 
would be extinct by 1980. In fact, many branches of
science seem destined to get ever closer to that point but never reach it.''

With online availability now of the scholarly literature it appears that
relatively fewer, rather than more,  papers are being cited and, by 
implication, read.  Evans \cite{evans} finds that:
``as more journal issues came online, the articles referenced tended 
to be more recent, fewer
journals and articles were cited, and more of those citations were to 
fewer journals and articles''.  Evans continues: 
``The forced browsing of print archives may have stretched scientists 
and scholars to anchor findings
deeply into past and present scholarship. Searching online is more 
efficient and following
hyperlinks quickly puts researchers in touch with prevailing opinion, 
but this may accelerate
consensus and narrow the range of findings and ideas built upon.''
Again notwithstanding the 34 million articles used in this study, it is 
clear that there are major divides between, say, mathematical 
methodology and large teams and consortia in the life and physical 
sciences.  

In a commentary on the Evans article \cite{evans}, Couzin \cite{couzin}
refers to ``herd behavior among authors'' in scholarly publishing.  
Couzin concludes by pointing to how this trend ``may lead to 
easier consensus and less active debate in academia''.  
 
I would draw the conclusion that mathematical
thinking -- if only because it lends itself poorly to the particular 
way that  ``prevailing opinion'' and acceleration of ``consensus'' are forced 
by how we now carry out research -- 
is of great importance for innovation and new thinking.  

The change in research and scholarly publishing has implications 
for book publishing.  Evans \cite{evans} notes this:  
``The move to online science appears to represent
one more step on the path initiated by the
much earlier shift from the contextualized
monograph, like Newton's {\em Principia} or
Darwin's {\em Origin of Species}, to the modern
research article. The {\em Principia} and {\em Origin}, each
produced over the course of more than a decade,
not only were engaged in current debates, but
wove their propositions into conversation with
astronomers, geometers, and naturalists from centuries
past. As 21st-century scientists and scholars
use online searching and hyperlinking to frame
and publish their arguments more efficiently, they
weave them into a more focused -- and more
narrow -- past and present.''

Undue focus and narrowness, and ``herd behavior'',  are at 
odds with the Hayashi and Benz\'ecri vision of science.  
Fortunately, this 
vision of science has not lost its sharp edge and its 
innovative potential for our times.

\section{Centrality of Data Analysis in Early Computer Science and 
Engineering}
\label{sect4}

\subsection{Data Stores: Prehistory of the Web}
\label{sect41}

Comprehensive and encyclopedic collection and interlinkage of 
data and information that is now typified by the web has a very long 
history.  Here I first point to just some of these antecedents that 
properly belong to the prehistory, well {\em avant la lettre}, of 
the web.

In the 12th century the web could maybe be typified by the work of 
John Tzetzes,
c.\ 1110--1180, who according to Browning \cite{browning}
was somewhat dysfunctional in his achievements:
`` ... His range was immense ... He had
a phenomenal memory ... philological commentaries on works of
classical Greek poetry ... works of scholarship in verse ... long,
allegorical commentaries ... encyclopedia of Greek mythology ...
long hexameter poems ... works of popularization ...
Tzetzes compiled a collection of his letters, as did many of his
contemporaries.  He then went on, however, and equipped it with a
gigantic commentary in nearly 13,000 lines of `political' verse,
which is a veritable encyclopedia of miscellaneous knowledge.
Later he went on to add the elements of a prose commentary on his
commentary.  The whole work conveys an impression of scholarship
without an object, of a powerful engine driving nothing.  Tzetzes
was in some ways a misfit and a failure in his own society.  Yet
his devotion of immense energy and erudition to a trivial end is
a feature found elsewhere in the literature of the twelfth century,
and points to a breakdown in the structure of Byzantine society and
Byzantine life, a growing discrepancy between ends and means.''

In modern times, the famous 1945 article \cite{vbush} by Vannevar Bush 
(1890--1974) set the scene in very clear terms for the web:
``Consider a future device for individual use, which is a sort of 
mechanized private file and library. It needs a name, and, to coin 
one at random, `memex' will do. A memex is a device in which an 
individual stores all his books, records, and communications, and 
which is mechanized so that it may be consulted with exceeding 
speed and flexibility. It is an enlarged intimate supplement to 
his memory.''

It is not widely recognized that Bush's famous essay was preceded
by an extensively developed plan by Belgian Paul Otlet, 1868--1944, 
especially in his book, {\em Trait\'e de Documentation} \cite{otlet}, 
published in Brussels in 1934.  

As described by him in the chapter entitled ``The preservation and 
international diffusion of thought: the microphotic book'' in 
\cite{rayward}, a clear view was presented (p. 208) of the physical, logical, 
and indeed socio-economical, layers necessary to support a prototype 
of the web: 
``By combining all the central offices discussed ...  one could create a 
``Document Super-Center.'' This would be in contact with national 
centers to which a 
country's principal offices of documentation and libraries would be 
linked to form stations in a universal network. ...
The books, 
articles and documents ... would be brought together in a great 
collection. Gradually a classified Microphotic Encyclopedia would 
be formed from them, 
the first step toward new microphotolibraries. All of these developments 
would be linked together to form a Universal Network of Documentation.''

\subsection{Between Classification and Search}

In the next three sections,
sections \ref{sect42}, \ref{sect43} and \ref{sect44}, 
 I progress through the physical and logical
layers supporting data analysis.  I detail just some of the work in the 
1960s and 1970s that involved the practicalities of data analysis.
Such work extended the work of Otlet and Bush.  While constituting 
just small building blocks in the massive edifice of what we now 
have by way of search and access to data and information, the contribution of 
underpinning theory and of skillful implementation that I discuss in these
sections should not be underestimated.  

Let me draw a line between the work of Bush and Otlet, which may
have been eclipsed for some decades, but which indicates nonetheless
that certain ideas were in the spirit of the times.  The disruptive
technology that came later with search engines like Google changed
the rules of the game, as the software industry often does.  Instead
of classifying and categorizing information, search and 
discovery were to prove fully sufficient.  Both classification and 
search were a legacy of the early years of exploratory data analysis
research.  Classification and search are two sides of the same coin.
Consider how the mainstay 
to this day of hierarchical clustering algorithms remains the nearest neighbor
chain and reciprocal nearest neighbor algorithms, developed in Benz\'ecri's
lab in the early 1980s \cite{murtagh85}.

The ftp protocol (file transfer protocol) was developed in the 1970s and 
took its definitive present form by 1985.  
Increasingly
wider and broader uptake of data and information access protocols 
was the order of the day by around 1990.  Archie, a search service
for ftp  was developed initially at the 
McGill University School of Computer Science in 1990.  The 
World Wide Web concept and http (hypertext transfer protocol) was
in development by Tim Berners-Lee at CERN by 1991.  In 1991 
a public version of Wide Area Information Servers (WAIS), 
invented by Brewster Kahle, was released by Thinking Machines Corporation 
WAIS was based on the Z39.50 and was a highly influential (certainly for 
me!) forerunner of web-wide information search and 
discovery. 
In April 1991 Gopher was released by the University of Minnesota Microcomputer,
Workstation and Networks Center. 
Initially the system was a university help service, a ``campus-wide 
document delivery system''.  In 1992, a search service for 
Gopher servers was developed under the name of Veronica, and 
released by the University of Nevada.
The University of Minnesota upset the burgeoning communities 
using wide area data and information search and discovery by
introducing licensing of Gopher.  This was just before the 
release of the Mosaic web browser, developed by Marc Andreessen,
an undergraduate student at the University of Illinois, Champaign.  
A contemporary vantage point on some of these developments is 
in my edited compilation \cite{iir}, which was finalized 
in the late summer of 1992.  

It may be noted too that the bibliographies in the Classification 
Literature Automated Search Services (see Appendix) were set up 
to be accessed through WAIS in the early 1990s.  

It is useful to have sketched out this subsequent evolution 
in data and information search and discovery because it 
constitutes one, but unquestionably an enormous, development 
rooted in earlier work on heterogeneous data collection and 
multivariate data analysis.  

\subsection{Environment and Context of Data Analysis}
\label{sect42}

I have noted that even in early times, the role of computational
capability was central (see sections \ref{sect2} and \ref{jpb68}).  

Describing early work with John Gower in the Statistics Department at 
Rothamsted Experimental Station in 1961, when Frank Yates was head of 
department, Gavin Ross reviewed data analysis as follows
\cite{ross}.  

``... we had several requests for classification jobs, mainly 
agricultural and biological at first, such as classification of 
nematode worms, bacterial strains, and soil profiles.  On this 
machine and its faster successor, the Ferranti Orion, we performed 
numerous jobs, for archaeologists, linguists, medical research 
laboratories, the Natural History Museum, ecologists, and even 
the Civil Service Department.  

On the Orion we could handle 600 units and 400 properties per unit, 
and we programmed several alternative methods of classification, 
ordination and identification, and graphical displays of the minimum 
spanning tree, dendrograms and data plots.  My colleague Roger Payne 
developed a suite of identification programs which was used to form a 
massive key to yeast strains.

The world of conventional multivariate statistics did not at first 
know how to view cluster analysis.  Classical discriminant analysis 
assumed random samples from multivariate normal populations.  Cluster 
analysis mixed discrete and continuous variables, was clearly not 
randomly sampled, and formed non-overlapping groups where multivariate 
normal populations would always overlap.  Nor was the choice of 
variables independent of the resulting classification, as Sneath 
had originally hoped, in the sense that if one performed enough 
tests on bacterial strains the proportion of matching results 
between two strains would reflect the proportion of common genetic 
information.  But we and our collaborators learnt a lot from these 
early endeavours.''

In establishing the Classification Society \cite{classsoc}, 
the interdisciplinary of the objectives was stressed: ``The foundation
of the society follows the holding of a Symposium, organized by 
Aslib on 6 April, 1962, entitled `Classification: an interdisciplinary
problem', at which it became clear that there are many aspects 
of classification common to such widely separated disciplines as 
biology, librarianship, soil science, and anthropology, and that 
opportunities for joint discussion of these aspects would be of 
value to all the disciplines concerned.''  

How far we have come can be seen in \cite{benzecri07} where 
target areas are sketched out that range over
analysis of voting and elections; jet algorithms for the Tevatron and 
Large Hadron Collider systems; gamma ray bursts; environment and 
climate management; sociology of religion; data mining in retail; 
speech recognition and analysis; sociology of natality -- analysis 
of trends and rates of births; and economics and finance -- 
industrial capital in Japan, financial data analysis in France, 
monetary and exchange rate analysis in the United States.  In all 
cases the underlying explanations are wanted, and not 
superficial displays or limited regression modeling.  

\subsection{Information Retrieval and Linguistics: Early Applications
of Data Analysis}
\label{sect43}

Roger Needham and Karen Sp\"arck Jones were two of 
the most influential figures in computing and the computational
sciences in the UK and worldwide.  

The work of Roger Needham, who died in February 2003,
 ranged over a wide swathe of computer
science. His early work at Cambridge in the 1950s included
cluster analysis and information retrieval.  In the 1960s, he
carried out pioneering work on computer architecture
and system software.  In the 1970s, his work involved
distributed computing.  In later decades, he devoted
considerable attention to security.

In the 1960s
he published on clustering and classification.  Information
retrieval was among the areas he contributed to.  Among his
early publications were:

\begin{enumerate}
\item ``Keywords and clumps'', Journal of Documentation, 20, 5--15, 1964.

\item ``Applications of the theory of clumps'', Mechanical Translation, 8,
113--127, 1965.

\item ``Automatic classification in linguistics'', The Statistician,
17, 45--54, 1967.

\item ``Automatic term classifications and retrieval'', Information
Storage and Retrieval, 4, 91--100, 1968.
\end{enumerate}

Needham, who was the husband of Sp\"arck Jones, 
 set up and became first director of Microsoft
Research in Cambridge in 1997.   

Karen Sp\"arck Jones died in April 2007.  Among early and 
influential publications on her side were the following.  

\begin{enumerate}

\item ``Experiments in semantic classification'', 
Mechanical Translation, 8, 97--112, 1965.

\item ``Some thoughts on classification for retrieval'', 
Journal of Documentation, 26, 89--101, 1970. (Reprinted 
in Journal of Documentation, 2005.)

\item With D.M. Jackson,
``The use of automatically-obtained keyword classifications 
for information retrieval'', Information Storage and Retrieval, 
5, 175--201, 1970.

\item {\em 
Automatic Keyword Classification for Information Retrieval}, 
Butterworths, 1971.

\end{enumerate}

Even in disciplines outside of formative or emergent computer science,
the centrality of data analysis algorithms is very clear from a scan of 
publications in earlier times.  A leader of classification and 
clustering research over many decades is James Rohlf (State University 
of New York).  As one 
among many examples, we note this work of his:  

F.J. Rohlf, Algorithm 76.
Hierarchical clustering using the
minimum spanning tree. Computer Journal, 
16, 93--95, 1973.

I will now turn attention to the early years of the Computer Journal. 

\subsection{Early Computer Journal}
\label{sect44}

A leader in early clustering developments and in information retrieval,
C.J. (Keith) van Rijsbergen (now Glasgow University) was Editor-in-Chief
of the Computer Journal from 1993 to 2000.   A few of his early papers
include the following.  

\begin{enumerate}
\item C.J. van Rijsbergen, ``A clustering algorithm'', Computer 
Journal, 13, 113--115, 1970.
\item N. Jardine and C.J. van Rijsbergen, ``The use of hierarchic
clustering in information retrieval'', Information Storage and 
Retrieval, 7, 217--240, 1971.
\item C.J. van Rijsbergen, ``Further experiments with hierarchic 
clustering in document retrieval'', Information Storage and Retrieval, 
10, 1--14, 1974.
\item C.J. van Rijsbergen, ``A theoretical basis for the use of 
co-occurrence data in information retrieval'', Journal of 
Documentation, 33, 106--119, 1977.

\end{enumerate}

From 2000 to 2007, I was
in this role as Editor-in-Chief of the Computer Journal.
I wrote in an editorial 
for the 50th Anniversary in 2007 the following:

``When I pick up older issues of the Computer Journal, I am struck 
by how interesting many of the articles still are.  Some articles 
are still very highly cited, such as Fletcher and Powell on gradient
descent.  Others, closer to my own heart, on clustering, data analysis, 
and information retrieval, 
by Lance and Williams, Robin Sibson, Jim Rohlf, Karen Sp\"arck Jones, 
Roger Needham, Keith van Rijsbergen, and others, to my mind established
the foundations of theory and practice that remain hugely important to this
day.  It is a pity that journal 
impact factors, which mean so much for our day to 
day research work, are based on publications in just two previous years.  
It is clear that new work may, or perhaps should, strike out to new shores,
and be unencumbered with past work.  But there is of course another 
also important view, that the consolidated literature is both vital and 
a well spring of current and future progress.   Both aspects are crucial, 
the `sleep walking' innovative element, to use Arthur Koestler's
\cite{koestler} 
characterization, and the consolidation element that is part and parcel of 
understanding.''  

The
very first issue of the Computer Journal in 1958 had 
articles by the following authors -- note the industrial 
research lab affiliations for the most part: 

\begin{enumerate}
\item S. Gill (Ferranti Ltd.), ``Parallel Programming'', pp. 2--10.
\item E.S. Page
\item D.T. Caminer (Leo Computers Ltd.)
\item R.A. Brooker (Computing Machine Laboratory, University of Manchester)
\item R.G. Dowse and H.W. Gearing (Business Group of the British Computer 
Society)
\item A. Gilmour (The English Electric Company Ltd.)
\item A.J. Barnard (Norwich Corporation)
\item R.A. Fairthorne (Royal Aircraft Establishment, Farnborough)
\item S.H. Hollngdale and M.M. Barritt (RAE as previous)
\end{enumerate}

Then from later issues I will note some articles that have very 
clear links with data analysis:

\begin{enumerate}
\item Vol. 1, No. 3, 1958, J.C. Gower, ``A note on an iterative 
method for root extraction'', 142--143.
\item Vol. 4, No. 1, 1961, M.A. Wright, ``Matching inquiries to 
an index'', 38--41.
\item Vol. 4, No. 2, 1961, had lots of articles on character
recognition.  
\item Vol. 4, No. 4, 1962, J.C. Gower, ``The handling of 
multiway tables on computers'', 280--286.
\item In Vol. 4, No. 4, and in Vol. 6, No. 1, there were 
articles on regression analysis.
\item Vol. 7, No. 2, 1964, D.B. Lloyd, ``Data retrieval'', 110--113.
\item Vol. 7, No. 3, 1964, M.J. Rose, ``Classification of a set of
elements'', 208--211. 
Abstract: ``The paper describes the use of a computer in some 
statistical experiments on weakly connected graphs.  The work forms
part of a statistical approach to some classification problems.''
\end{enumerate}

\section{Conclusions}

In this article I have focused on early developments in the 
data mining or unsupervised view of data analysis.  Some of those
I have referred to, e.g.\ Vannevar Bush and Paul Otlet, became 
obscured or even eclipsed for a while.  
It is clear however that 
undisputable progress in the longer term may seem to 
develop in fits and starts when seen at finer temporal scales. 
(This is quite commonplace.  In literature, see how the centenary of 
Goethe's birth in 1848, following his death on 22 March 1832, 
passed unnoticed.  Goethe did not come to the fore until the 
1870s.)  

What I am dealing with therefore in exploratory, 
heuristic and multivariate data analysis has led me to a sketch of
the evolving spirit of the times.  This sketch has taken in 
the evolution 
of various strands in academic disciplines, scholarly 
research areas, and commercial, industrial and economic 
sectors.  

I have observed how the seeds of the present -- in fact, 
remarkably good likenesses -- were often available in the 
period up to 1985 that is mainly at issue in this article.  This includes the 
link between scholarly activity and economic and commercial 
exploitation.  It includes various aspects of the PhD degree.  

The consequences of the data mining and related exploratory 
multivariate data analysis work overviewed in this article 
have been enormous.  Nowhere have
their effects been greater than in current search engine technologies.  
Also wide swathes of database management, language engineering,
and multimedia data and digital information handling, are all
directly related to the pioneering work described in this 
article.  

In section \ref{sect41} I looked at how early exploratory data
analysis had come to play a central role in our computing
infrastructure.  An interesting view has been offered by \cite{wired},
finding that all of science too has been usurped by exploratory data
analysis, principally through Google's search facilities.  Let us
look at this argument with an extended quotation from \cite{wired}.

`` `All models are wrong, but some are useful.'
So proclaimed statistician George Box 30 years ago, and he was right.
... Until now. ...

At the petabyte scale, information is not a matter of simple three- and
four-dimensional taxonomy and order but of dimensionally agnostic
statistics. It calls for an entirely different approach, one that
requires us to lose the tether of data as something that can be
visualized in its totality. It forces us to view data mathematically
first and establish a context for it later. ...

Speaking at the O'Reilly Emerging Technology Conference this past March
[2008],
Peter Norvig, Google's research director, offered an update to George
Box's maxim: `All models are wrong, and increasingly you can succeed
without them.' ...

This is a world where massive amounts of data and applied mathematics
replace every other tool that might be brought to bear. Out with every
theory of human behavior, from linguistics to sociology. Forget taxonomy,
ontology, and psychology. Who knows why people do what they do? The point
is they do it, and we can track and measure it with unprecedented
fidelity. With enough data, the numbers speak for themselves.

The big target here isn't advertising, though. It's science. The
scientific method is built around testable hypotheses. These models,
for the most part, are systems visualized in the minds of scientists.
The models are then tested, and experiments confirm or falsify
theoretical models of how the world works. This is the way science
has worked for hundreds of years. ...

There is now a better way. Petabytes allow us to say: `Correlation
is enough.' We can stop looking for models. We can analyze the data
without hypotheses about what it might show. We can throw the numbers
into the biggest computing clusters the world has ever seen and let
statistical algorithms find patterns where science cannot.''

This interesting view, inspired by our contemporary search engine
technology, is provocative.  The author maintains 
that: 
``Correlation supersedes causation, and
science can advance even without coherent models, unified theories,
or really any mechanistic explanation at all.''  

No, in my view, the sciences and humanities 
are not to be consigned to any dustbin of history -- far from it.

As I wrote in \cite{deleeuw}, a partnership is needed rather than
dominance of one view or another.
``Data analysts have far too often just assumed
the potential for extracting meaning from the given data, {\em telles 
quelles}.  The statistician's way to address the problem works well sometimes
but has its limits: some one or more of a finite number of
stochastic models (often handled with the verve and adroitness of a maestro)
form the basis of the analysis.  The
statistician's toolbox (or surgical equipment, if you wish)
can be enormously useful in practice.  But the statistician plays
second fiddle to the observational scientist or theoretician who really
makes his or her mark on the discovery.  This is not fair.

Without exploring the encoding that makes up primary data we know very,
very little.  (As examples, we have the DNA codes of the human or any
animal; discreteness at Planck scales and in one vista of the quantum
universe; and we still have to find the proper encoding to understand
consciousness.)   ... [Through correspondence analysis there]
is the possibility opened up for the data
analyst, through the data encoding question, to be a partner, hand in
hand, in the process of  primary discovery.''

\section*{Appendix: Sources for Early Work}

\begin{itemize}

\item Classification Literature Automated Search Service, a CD 
distributed currently with the first issue each year of the Journal 
of Classification.  See http://www.classification-society.org/csna

The following books have been scanned and are available in their 
entirety on the CD.  

\begin{enumerate}
\item Algorithms for Clustering Data (1988), AK Jain and RC Dubes
\item Automatische Klassifikation (1974), HH Bock
\item Classification et Analyse Ordinale des Donn\'ees (1981), IC Lerman
\item Clustering Algorithms (1975), JA Hartigan
\item Information Retrieval (1979, 2nd ed.), CJ van Rijsbergen
\item Multidimensional Clustering Algorithms (1985), F Murtagh 
\item Principles of Numerical Taxonomy (1963), RR Sokal and PHA Sneath
\item Numerical Taxonomy: the Principles and Practice of Numerical 
Classification (1973), PHA Sneath and RR Sokal
\end{enumerate}

\item {\em Les Cahiers de l'Analyse des Donn\'ees} was the journal 
of Benz\'ecri's lab from 1975 up to 1997, with four issues per year.  
Scanning of all issues has started, working chronologically backwards
with thus far 1994--1997 covered.
See http://thames.cs.rhul.ac.uk/$\sim$fionn/CAD

\item Some texts by Jean-Paul Benz\'ecri and 
Fran\c{c}oise Benz\'ecri-Leroy,
published between 1954 and 1971, are available at http://www.numdam.org
(use e.g.\ ``benz\'ecri'' as a search term).

\end{itemize}


\begin{thebibliography}{99}

\bibitem{aczel}
A.D. Aczel, {\em The Artist and the Mathematician: The Story of 
Nicolas Bourbaki, the Genius Mathematician Who Never Existed},
High Stakes, 2006. 

\bibitem{adler}
 R Adler, J Ewing, P Taylor, {\em Citation Statistics}, A report from the
 International Mathematical Union (IMU) in cooperation with the International
 Council of Industrial and Applied Mathematics (ICIAM) and the
 Institute of Mathematical Statistics (IMS),
 Joint Committee on Quantitative Assessment of Research, 11 June 2008

\bibitem{wired}
C. Anderson, ``The end of theory: the data deluge makes the scientific
method obsolete'', {\em Wired Magazine}, 16 July 2008,
http://www.wired.com/science/discoveries/magazine/16-07/pb\_theory

\bibitem{badioux}
A. Badiou, {\em Theoretical Writings}, edited and translated by
Ray Brassier and Alberto Toscana, Continuum, 2004.

\bibitem{benjamin}
W. Benjamin, ``Problems in the sociology of language'', in 
{\em Walter Benjamin, Selected Writings: 1935--1938 Vol. 3},
Harvard University Press, 2002.  

\bibitem{benzecri82}
J.-P. Benz\'ecri, {\em Histoire et Pr\'ehistoire de l'Analyse des
Donn\'ees}, Dunod, 1982. 

\bibitem{jpbavenir}
J.P. Benz\'ecri, ``L'avenir de l'analyse des donn\'ees'',
{\em Behaviormetrika}, 10, 1--11, 1983.
Accessible from: www.correspondances.info

\bibitem{jpb0}
J.-P. Benz\'ecri, ``Foreword'', in \cite{murtagh05}, 2005.

\bibitem{benzecri07} 
J.-P. Benz\'ecri, ``Si j'avais un laboratoire...'', 11 pp., 2007.
Scanned copy at www.correspondances.info  

\bibitem{browning}
R. Browning, {\em The Byzantine Empire}, Catholic University of 
America Press, 1992.

\bibitem{vbush}
V. Bush, ``As we may think'', {\em Atlantic Monthly}, July 1945.
http://www.theatlantic.com/doc/194507/bush

\bibitem{chomsky}
N. Chomsky, {\em Syntactic Structures}, 2nd edn., Walter de Gruyter,
2002. 

\bibitem{clark} 
W. Clark, {\em Academic Charisma and the Origins of the Research 
University}, Chicago University Press, 2006. 

\bibitem{classsoc}
The Classification Society, Record of the inaugural meeting, held at the
offices of Aslib, 3 Belgrave Square, London S.W. 1 at 2.30 p.m. on 
Friday, 17 April, 1964. \\
 http://thames.cs.rhul.ac.uk/$\sim$fionn/classification-society/ClassSoc1964.pdf

\bibitem{couzin}
J. Couzin, ``Survey finds citations growing narrower as journals move 
online'', {\em Science}, 321, 329, 2008.  

\bibitem{derham}
C. de Rham, ``La classification hi\'erarchique ascendante selon la 
m\'ethode des voisins r\'eciproques'', {\em Les Cahiers de l'Analyse
des Donn\'ees}, V, 135--144, 1980.  

\bibitem{dijkstra}
E.W. Dijkstra, ``The humble programmer'', ACM Turing Lecture, 1972,
{\em Communications of the ACM}, 15 (10), 859--866, 1972.

\bibitem{evans}
J.A. Evans, ``Electronic publishing and the narrowing of science and 
scholarship'', {\em Science}, 321, 395--399, 2008.

\bibitem{foucault} 
M. Foucault, {\em Surveiller et Punir: Naissance de la Prison},
Gallimard, 1975. 

\bibitem{galileo}
Galileo Galilei, Il Saggiatore (The Assayer) in {\em Opere}, vol. 6, p. 197,
translation by Julian Barbour.  Cited on p. 659 of 
N. Stephenson, {\em Quicksilver, The Baroque Cycle, Vol. 1}, William 
Morrow, 2003. 

\bibitem{guedon}
J.-C. Gu\'edon, {\em In Oldenburg's Long Shadow: Librarians, Research 
Scientists, Publishers, and the Control of Scientific Publishing},
Association of Research Libraries, 2001, 70 pp.,
http://www.arl.org/resources/pubs/mmproceedings/138guedon.shtml

\bibitem{hauser}
M.D. Hauser and T. Bever, ``A biolinguistic agenda'', {\em Science}, 322, 
1057--1059, 2008.

\bibitem{hayashi0}
C. Hayashi, ``On the prediction of phenomena from qualitative data
and the quantification of qualitative data from the
mathematico-statistical point of view'', {\em Annals of the 
Institute of Statistical Mathematics}, 3, 69--98, 1952.
Available at: 
http://www.ism.ac.jp/editsec/aism/pdf/003\_2\_0069.pdf

\bibitem{hayashi}
C. Hayashi, ``Multidimensional quantification I'',
{\em Proceedings of the Japan Academy}, 30, 61--65, 1954.
Available at: http://www.ism.ac.jp/editsec/aism/pdf/005\_2\_0121.pdf

\bibitem{iir}
A. Heck and F. Murtagh, Eds., {\em Intelligent Information 
Retrieval: The Case of Astronomy and Related Space Science},
Kluwer, 1993.  

\bibitem{johns}
A. Johns, {\em The Nature of the Book: Print and Knowledge in the 
Making}, University of Chicago Press, 1998. 

\bibitem{jones}
D.H.P. Jones, ``Was the Carte du Ciel an obstruction to the 
development of astrophysics in Europe?'', in A. Heck, Ed., 
{\em Information Handling in Astronomy -- Historical Vistas}, 
Springer, 2002.  

\bibitem{juan1}
J. Juan, ``Le programme HIVOR de classification ascendante hi\'erarchique
selon les voisins r\'eciproques et le crit\`ere de la variance'',
{\em Les Cahiers de l'Analyse des Donn\'ees}, VII, 173--184, 1982. 

\bibitem{juan2}
J. Juan, ``Programme de classification hi\'erarchique par l'algorithme de la 
recherche en cha\^{\i}ne des voisins r\'eciproques'', {\em Les Cahiers
de l'Analyse des Donn\'ees}, VII, 219--225, 1982.  

\bibitem{koestler}
A. Koestler, {\em 
The Sleepwalkers: A History of Man's Changing Vision of 
the Universe}, Penguin, 1989.  (Originally published 1959.)

\bibitem{mathgen}
Mathematics Genealogy, www.genealogy.ams.org (127,901 entries as of
15 November 2008). 

\bibitem{miller}
G. Miller, ``Growing pains for fMRI'', {\em Science}, 320, 1412--1414, 
13 June 2008.

\bibitem{mur83a}
F. Murtagh, ``A survey of recent advances in hierarchical clustering 
algorithms'', {\em The Computer Journal}, 26, 354--359, 1983.

\bibitem{mur83b}
F. Murtagh, ``Expected-time complexity results for hierarchic clustering 
algorithms which use cluster centres'', {\em Information Processing 
Letters}, 
16, 237--241, 1983.

\bibitem{mur84}
F. Murtagh, ``Complexities of hierarchic clustering algorithms: state of 
the art'', {\em Computational Statistics Quarterly}, 1, 101--113, 1984.

\bibitem{murtagh85}
F. Murtagh, {\em Multidimensional Clustering Algorithms}, 
Physica-Verlag, W\"urzburg, 1985. 

\bibitem{murtagh05}
F. Murtagh, {\em Correspondence Analysis and Data Coding with Java and R}, 
Chapman and Hall/CRC, 2005. 

\bibitem{mursw}
F. Murtagh, ``Multivariate data analysis software and resources'',
http://astro.u-strasbg.fr/$\sim$fmurtagh/mda-sw

\bibitem{deleeuw}
F. Murtagh, ``Reply to: Review by Jan de Leeuw of Correspondence
Analysis and Data Coding with Java and R, F. Murtagh,
Chapman and Hall/CRC, 2005''. Journal of Statistical Software,
Vol. 14.
http://www.correspondances.info/reply-to-Jan-de-Leeuw.pdf

\bibitem{murtagh08}
F. Murtagh, ``Between the information economy and 
student recruitment: present conjuncture and future prospects'', 
{\em Upgrade: The European Journal for the Informatics Professional},
forthcoming, 2008.  

\bibitem{nihoa}
National Institutes of Health Public Access, 
http://publicaccess.nih.gov, 2008.
Citation from: ``The NIH Public Access Policy implements Division G, Title II, 
Section 218 of PL 110-161 (Consolidated Appropriations Act, 2008).''

\bibitem{otlet}
P. Otlet, {\em Trait\'e de Documentation}, Brussels, 1934. 
https://archive.ugent.be/handle/1854/5612

\bibitem{rayward}
W.B. Rayward, {\em International Organisation and Dissemination of 
Knowledge: Selected Essays of Paul Otlet}, Elsevier, 1990.
http://www.archive.org/details/internationalorg00otle

\bibitem{ross}
G. Ross, ``Earlier days of computer classification'', Vote of thanks, 
Inaugural Lecture, F. Murtagh, ``Thinking ultrametrically: understanding
massive data sets and navigating information spaces'', Royal Holloway, 
University of London, 22 Feb.\ 2007, 
http://thames.cs.rhul.ac.uk/$\sim$fionn/inaugural

\bibitem{stephenson}
N. Stephenson, {\em The System of the World, The Baroque Cycle, Vol. 3}, 
Arrow Books, 2005. 

\bibitem{suzanne}
B. Suzanne, ``Frequently asked questions about Plato'', 2004, 
http://plato-dialogues.org/faq/faq009.htm

\bibitem{swade}
D. Swade, {\em The Cogwheel Brain: Charles Babbage and the Quest to 
Build the First Computer}, Little, 2000.  

\bibitem{wuester}
Eugen W\"uster, {\em Internationale Sprachnormung
in der Technik, besonders in der Elektrotechnik}, Bern, 1931.

\bibitem{whitfield}
J. Whitfield, ``Collaboration: group theory'', Nature News items,
{\em Nature}, 455, 720--723, 2008.  

\bibitem{yourgrau}
P. Yourgrau, {\em A World Without Time: The Forgotten Legacy of 
G\"odel and Einstein}, Allen Lane, 2005.  

\end{thebibliography}
\end{document}